\documentclass[a4paper,11pt]{article}

\pdfoutput=1

\usepackage[a4paper,left=2.73cm,right=2.7cm,top=3cm,bottom=3.5cm]{geometry}

\usepackage[T1]{fontenc} 
\usepackage{graphicx}
\usepackage{epsfig}
\usepackage{amssymb}
\usepackage{amsfonts}
\usepackage{amsmath,euscript,array,mathrsfs}
\usepackage{bbold}
\usepackage{epsf}
\usepackage{slashed}
\usepackage{comment}
\usepackage{colortbl}

\usepackage{caption,subcaption,wrapfig}
\usepackage[colorlinks=true,linktocpage=true,linkcolor=blue,citecolor=blue]{hyperref}

\usepackage{tikz}
\usetikzlibrary{decorations.markings,decorations.pathmorphing}

\newcommand{\be}{\begin{equation}}
\newcommand{\ee}{\end{equation}}
\newcommand{\beqs}{\begin{eqnarray}}
\newcommand{\eeqs}{\end{eqnarray}}
\newcommand{\parent}[1]{\left(#1\right)}

\newcommand{\dd}{\mathrm{d}}

\def\cO{{\cal O}}

\begin{document}
 \begin{titlepage}

\thispagestyle{empty}

\begin{flushright}
\hfill{ICCUB-19-015}
\end{flushright}

\vspace{40pt}  
	 
\begin{center}

{\huge \textbf{Holographic Complex Conformal Field Theories}}

\vspace{30pt}
		
{\large \bf Ant\'on F. Faedo,$^{1}$   Carlos Hoyos,$^{2,\,3}$   \\ [1mm]
David Mateos$^{1,\,4}$  and Javier G. Subils$^{1}$}

\vspace{25pt}

{\normalsize  $^{1}$ Departament de F\'\i sica Qu\'antica i Astrof\'\i sica and Institut de Ci\`encies del Cosmos (ICC),\\  Universitat de Barcelona, Mart\'\i\  i Franqu\`es 1, ES-08028, Barcelona, Spain.}\\
\vspace{15pt}
{ $^{2}$Dept. of Physics, Universidad de Oviedo, \\ Federico Garc\'ia Lorca 18, ES-33007, Oviedo, Spain.}\\
\vspace{15pt}
{ $^{3}$Instituto Universitario de Ciencias y Tecnolog\'{\i}as Espaciales de Asturias (ICTEA), \\ Calle de la Independencia, 13, ES-33004, Oviedo, Spain.}\\
\vspace{15pt}
{ $^{4}$Instituci\'o Catalana de Recerca i Estudis Avan\c cats (ICREA), \\ Passeig Llu\'\i s Companys 23, ES-08010, Barcelona, Spain.\\ 
}


\vspace{40pt}
				
\abstract{The loss of criticality in the form of weak first-order transitions or the end of the conformal window in gauge theories can be described as the merging of two fixed points that move to complex values of the couplings. When the complex fixed points are close to the real axis, the system typically exhibits walking behavior with Miransky (or Berezinsky--Kosterlitz--Thouless) scaling. We present a novel realization of these phenomena at strong coupling by means of the gauge/gravity duality, and give evidence for the conjectured existence of complex conformal field theories at the fixed points.}

\end{center}

\end{titlepage}

\tableofcontents

\hrulefill
\vspace{10pt}

\section{Introduction}

Fixed-point annihilation (FPA) is an interesting phenomenon in which two fixed points (FPs) of the renormalization group (RG) flow merge and disappear as some parameter is varied. In the context of phase transitions and critical phenomena it is associated to a change from continuous to weak 
first-order transitions. Examples include the superconducting transition in the Abelian Higgs model \cite{Halperin:1973jh,Ihrig:2019kfv}, the related N\'eel-valence bond-solid transition in antiferromagnets \cite{Senthil:2004,Nahum:2015jya,Wang:2017txt,Serna:2018tct}, the ferromagnetic transition in the Potts model \cite{Nienhuis:1979mb,Nauenberg:1980nv,Gorbenko:2018dtm}, metal-Mott insulator transitions \cite{Herbut:2014lfa} and six-dimensional $O(N)$ models \cite{Fei:2014xta,Gracey:2018khg}. FPA has also been associated to the boundaries of the conformal window in gauge theories with flavors, both in $(2+1)$-dimensional quantum electrodynamics \cite{Appelquist:1988sr,Kubota:2001kk,Kaveh:2004qa,Herbut:2016ide} and in non-Abelian gauge theories in $3+1$ dimensions \cite{Gies:2005as,Pomoni:2008de,Kaplan:2009kr,Antipin:2012kc,Hansen:2017pwe}.

More generally, FPA has been proposed as a natural mechanism to produce ``walking behavior'' in gauge theories \cite{Gorbenko:2018ncu}. The idea is that, just after the merging, the critical points leave a footprint in the form of approximate scale invariance over a large range of scales. Typically the range of the walking region increases exponentially as parameters are tuned to the merging point, following  Miransky (or Berezinsky--Kosterlitz--Thouless) scaling \cite{Miransky:1984ef,Berezinsky:1970fr,Kosterlitz:1973xp}. This behavior can be explained by continuing the theory to complex values of the couplings, so that the annihilation is understood as a migration of the FPs to the complex plane after the merger \cite{Kaplan:2009kr}. Their effect on the RG flow is noticeable as long as they remain close to the real axis. It has been recently conjectured  \cite{Gorbenko:2018ncu} that a non-unitary, complex conformal field theory (CCFT) exists at each of the two complex fixed points (CFPs), so that the properties of the theory in the walking region can be derived from perturbations of the CCFTs. Each CCFT has a complex spectrum of operators that is the conjugate of its companion's, implying that CFPs should always come in pairs.

Although FPA and CFPs are expected to exist generically, their study has been mostly limited to weakly coupled theories (see e.g.~\cite{Sieg:2016vap,Grabner:2017pgm,Pittelli:2019ceq,Benini:2019dfy} for recent examples), as their identification requires computing the beta functions for the different couplings in the theory, a task that  often can only be done via perturbation theory. In this paper we will construct a simple holographic model that realizes FPA and CFPs, thus showing that these phenomena  can also occur at strong coupling. In addition, our analysis provides non-perturbative evidence that CFPs have the conjectured properties of CCFTs regarding the spectrum of local operators.

\section{Fixed-point annihilation and complex CFTs}
\label{sec:FPAandCCFT}

Consider  a system with a dimensionless coupling $g$ whose $\beta$-function depends on an external parameter $\alpha$ in such a way that, for $\alpha\simeq \alpha_*$, 
\begin{equation}\label{eq:beta}
\beta\left(g\right)\,\simeq \,\left(\alpha-\alpha_*\right)-\left(g-g_*\right)^2.
\end{equation} 
We will see an explicit  example in Sec.~\ref{sec:CFPandRGflows}. 
If $\alpha>\alpha_*$, discarding higher-order terms, the $\beta$-function vanishes at two values  
\begin{equation}
g_{\pm}\,=\,g_*\pm\sqrt{\alpha-\alpha_*}\,.
\end{equation} 
To make sure that the theory is well defined in the far ultraviolet (UV)  we may imagine that $\beta$ has another zero at some $g_{\infty}<g_-$.
Decreasing the control parameter $\alpha$ the FPs $g_\pm$ approach each other until they merge at \mbox{$\alpha=\alpha_*$}. If we decrease $\alpha$ further, $\beta(g)$ loses these (real) zeroes and the theory ceases to have a (real) conformal phase in the infrared (IR). However, for \mbox{$|\alpha-\alpha_*|$} sufficiently small, $\beta(g)$ has CFPs close to the real axis at $g_\pm=g_*\pm i \sqrt{\alpha_*-\alpha}$. In this regime the theory  exhibits approximate scale invariance between UV and IR  scales $\mu_{\text{\tiny UV}}$ and $\mu_{\text{\tiny IR}}$ defined by the values of the coupling $g_{\text{\tiny UV}}\lesssim g_*\lesssim g_{\text{\tiny IR}}$. The ratio between these two scales becomes exponentially large as $\alpha$ approaches $\alpha_*$ and shows the characteristic Miransky scaling
\begin{equation}\label{eq:BKT}
\log\frac{\mu_{\text{\tiny UV}}}{\mu_{\text{\tiny IR}}}\,=\,\int_{g_{\text{\tiny IR}}}^{g_{\text{\tiny UV}}}\frac{\dd g}{\beta(g)}\simeq \frac{\pi}{\sqrt{\alpha_*-\alpha}}\,,
\end{equation}
where we assumed that $\left|g_{\text{\tiny IR,UV}}-g_*\right|\gg\sqrt{\alpha_*-\alpha}$. Thus if $|\alpha-\alpha_*|$ is small then the RG flow is slow in a large energy range, hence the term ``walking'' flow. 

The scenario conjectured in \cite{Gorbenko:2018ncu, Gorbenko:2018dtm} is that CFPs correspond to pairs of non-unitary CCFTs that control the walking flow, which passes precisely in between them. Each CCFT should have operators of complex dimensions in the spectrum that are not matched by other operators with complex conjugate dimension in the same theory. Instead, the missing operators of complex conjugate dimensions in one CCFT should be part of the the spectrum of the companion CCFT. Thus in this sense the two CCFTs are complex conjugate of one another. 
 
An important case is the operator associated to the coupling $g$ itself, whose complex dimensions at each CFP, 
\begin{equation}\label{eq:Delta}
\Delta_{\pm}\,=\,d+\beta^\prime(g_{\pm})\,\simeq\,d\mp 2i\,\sqrt{\alpha_*-\alpha}\,,
\end{equation}
with $d$ the spacetime dimension, 
are indeed complex conjugates of one another. Moreover, this operator is close to marginality when $\alpha \lesssim \alpha_*$, with the leading deviation being imaginary and small. Using this, the hierarchy \eqref{eq:BKT} can be rewritten as 
\be
\label{Miransky}
\log \frac{\mu_{\text{\tiny UV}}}{\mu_{\text{\tiny IR}}} \simeq 
\frac{2\pi}{|\operatorname{Im}{\Delta}|} \,.
\ee

\section{Holographic realization}
\label{sec:holorealization}

Previous holographic realizations of walking behavior with Miransky-like scaling in gravity duals \cite{Kaplan:2009kr,Jensen:2010ga,Iqbal:2010eh,Pomarol:2019aae,Jarvinen:2011qe} are based on flows where the mass of a scalar field on the gravity side violates the Breitenlohner-Freedman (BF) bound \footnote{See however \cite{Alanen:2010tg,Alanen:2011hh} for a holographic model whose $\beta$-function shares some properties with the lower one in our Fig.~\ref{fig:beta}.}. In these cases there is a real FP that becomes dynamically unstable but no CFPs have been identified. We will present a different  construction in which both FPA and the resulting CFPs are explicitly realized. 

\begin{figure}[t]
	\begin{center}
			\includegraphics[width=0.45\textwidth,height=0.3\textwidth]{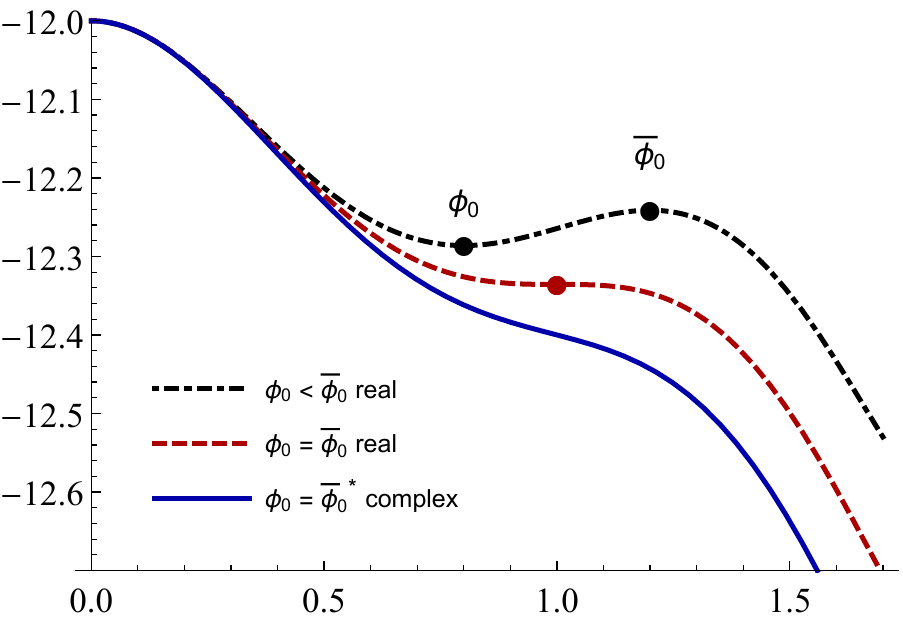} 
			\put(-200,142){ $L^2 V(\phi)$}
			\put(-7,0){$ \phi$}
		\caption{\small  Potential  of our model for the cases 
		$\{\phi_0=0.8, \overline{\phi}_0=1.2\}$ (top curve), 
		$\{\phi_0=\overline{\phi}_0=1\}$ (middle curve) and 
		$\{\phi_0= \overline{\phi}_0^*=1+0.2i\}$ (bottom curve).}
		\label{fig:pot}
	\end{center}
\end{figure}

Couplings of the gauge theory are holographically dual to fields on the gravity side. For simplicity we focus on a single coupling dual to a scalar field. The action on the gravity side is thus
\begin{equation}\label{eq:action}
S\,=\,\frac{1}{2\kappa^2}\int d^{d+1}x\sqrt{-g}\left(R-\frac12 (\partial\phi)^2-V(\phi)\right)\,.
\end{equation}
For each critical point $\phi_c$ of the potential with $V(\phi_c)<0$ there is an anti-de Sitter (AdS) solution with its corresponding $d$-dimensional CFT dual. We choose to write $V$ in terms of a (fake) superpotential $W$ through the usual relation
\begin{equation}\label{W}
V\,=\,\left(d-1\right)\left[2\left(d-1\right)\left(\frac{\dd W}{\dd\phi}\right)^2-d\,W^2\right] \,.
\end{equation} 
The only reason for this is to simplify the presentation. In particular, this choice implies nothing regarding the possible presence of supersymmetry in the system. Critical points of $W$ are also critical points of $V$ (but not viceversa). Since we wish to model three FPs we take a superpotential with derivative
\begin{equation}\label{dW}
\frac{\dd W}{\dd\phi}\,=\,\frac{W_0}{L}\,\phi\left(\phi-\phi_0\right)\left(\phi-\overline{\phi}_0\right).
\end{equation}
The resulting potential is shown in Fig.~\ref{fig:pot}.
The UV FP dual to the AdS solution at $\phi=0$ is the analog of the FP at \mbox{$g=g_{\infty}$} in Sec.~\ref{sec:FPAandCCFT}. The constants $\phi_0$ and $\overline{\phi}_0$ are parameters of the model analogous to $\alpha$. If both $\phi_0$ and $\overline{\phi}_0$ are real then there are two additional real FPs at \mbox{$\phi=\phi_0$} and $\phi=\overline{\phi}_0$, in analogy with  $g=g_\pm$ in Sec.~\ref{sec:FPAandCCFT}. 
When $\phi_0=\overline{\phi}_0$ these FPs merge into a single one. If 
$\phi_0$ and $\overline{\phi}_0$ become complex then they must be conjugate to one another since $W$ must be real for real $\phi$. In this case the potential looses two real critical points, giving a holographic realization of FPA \footnote{The potential determined from \eqref{dW} may have additional FPs that will  play no role in our discussion.}.

At the UV FP the AdS radius is fixed by the integration constant $W(0)=1/L$, while the dimension $\Delta_{\text{\tiny UV}}$ of the operator $\mathcal{O}$ dual to $\phi$ is determined by $W_0= \delta_{\text{\tiny UV}}/[2(d-1)\phi_0\overline{\phi}_0]>0$. There are two possible choices depending on whether the flow is triggered by a source for $\mathcal{O}$, in which case $\delta_{\text{\tiny UV}}=  d- \Delta_{\text{\tiny UV}}$, or by a non-zero expectation value for $\mathcal{O}$, in which case $ \delta_{\text{\tiny UV}}= \Delta_{\text{\tiny UV}}$.

Expanding the superpotential around $\phi=\phi_0$ we get
\begin{equation}\label{expansion}
W(\phi) =\frac{1}{L_0}+\frac{\delta_0}{4(d-1)L_0}(\phi-\phi_0)^2+{O}(\phi-\phi_0)^3\,,
\end{equation}
where
\begin{equation}\label{eq:L0W}
L_0 =L \left[1+\frac{W_0}{6}\left(\phi_0^3\overline{\phi}_0-\frac{\phi_0^4}{2} \right) \right]^{-1}
\end{equation}
is the AdS radius at $\phi=\phi_0$ and 
\begin{equation} \label{eq:D0W}
\delta_0 = 2(d-1) W_0\frac{L_0}{L}\phi_0(\phi_0-\overline{\phi}_0)=\delta_{\text{\tiny UV}}\frac{L_0}{L}\frac{(\phi_0-\overline{\phi}_0)}{\overline{\phi}_0}\,.
\end{equation}
The expansion around $\overline{\phi}_0$ gives analogous results with the replacements $\{ \phi_0, \overline{\phi}_0, L_0, \delta_0 \}\to \{ \overline{\phi}_0,\phi_0,  \overline{L}_0, \overline{\delta}_0 \}$. Assuming \mbox{$0<\phi_0 \lesssim \overline{\phi}_0$} we have that 
\be
\delta_0<0 \,, \qquad 0 < \overline{\delta}_0 < \frac{d}{2}-1 \,.
\ee
In this case $\overline{\phi}_0$ corresponds to an UV FP deformed by a relevant scalar operator of dimension \mbox{$\overline{\Delta}_0=d-\overline{\delta}_0$}, whereas $\phi_0$ corresponds to an IR FP deformed by an irrelevant  scalar operator of dimension  
$\Delta_0=d-\delta_0$. When $\phi_0=\overline{\phi}_0$ the two points merge and the dual operator becomes marginal.

\subsection{Complex FPs and RG flows}
\label{sec:CFPandRGflows}

\begin{figure}[t]
	\begin{center}
			\includegraphics[width=0.45\textwidth,height=0.3\textwidth]{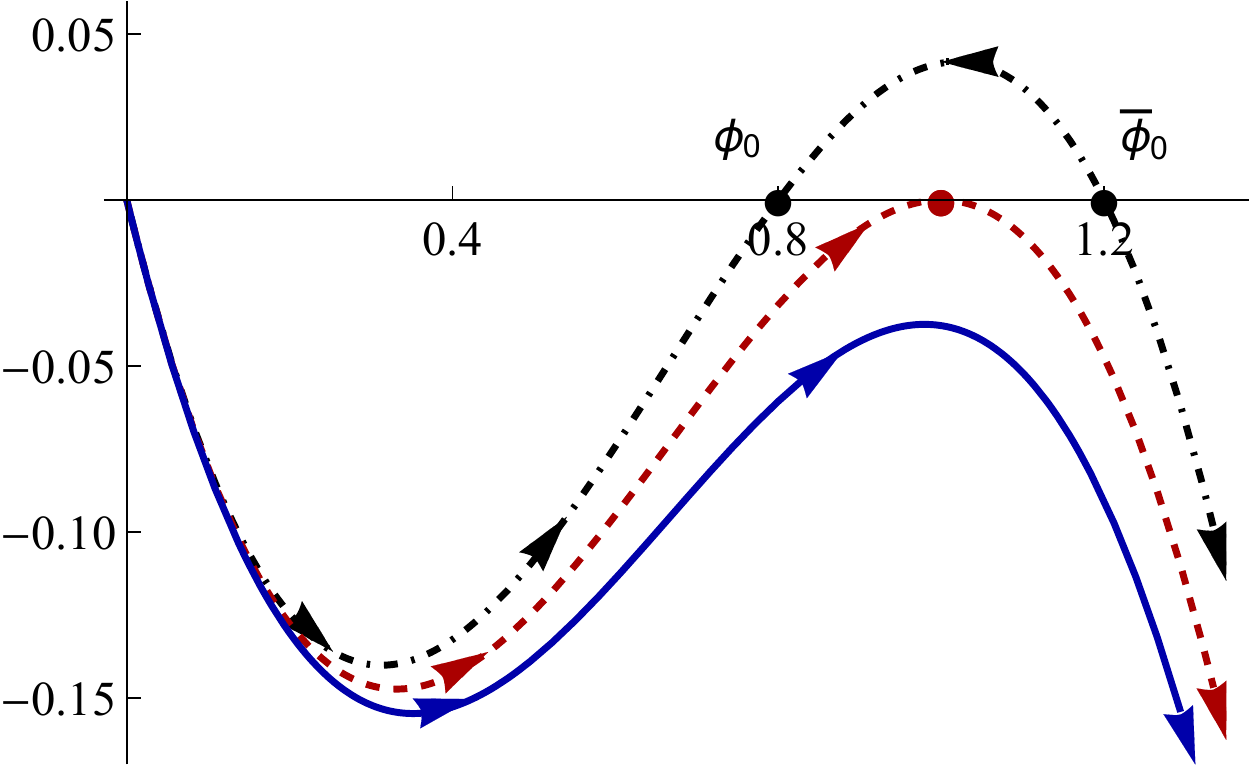} 
			\put(-175,120){$\beta(\phi)$}
			\put(-7,90){$ \phi$}
		\caption{\small  $\beta$-functions associated to each of the three potentials of Fig.~\ref{fig:pot}, as defined in \eqref{eq:betaphi}. The arrows indicate the direction of the RG flow from the UV to the IR. }
		\label{fig:beta}
	\end{center}
\end{figure}

In the so-called ``domain wall'' coordinates in which the metric takes the form 
\begin{equation}
\dd s_{d+1}^2\,=\,g_{MN}dx^M dx^N\,=\, e^{2A\left(\rho\right)}\dd x_{1,d-1}^2+\dd\rho^2
\end{equation}
the solution is determined by the  equations
\begin{equation}\label{BPS}
\frac{\dd A}{\dd\rho}\,=\, W\,,\qquad\qquad \frac{\dd \phi}{\dd\rho}\,=\,-2\left(d-1\right)\frac{\dd W}{\dd\phi}\,.
\end{equation}
In these coordinates the metric is foliated by copies of $d$-dimensional Minkowski space with scale factor $e^{A(\rho)}$, which is therefore interpreted as dual to the RG scale in the gauge theory. Similarly, the scalar field $\phi=\phi(\rho)$ is dual to a running coupling constant in a particular scheme, whose  $\beta$-function is therefore 
(see e.g.~\cite{Anselmi:2000fu})
\begin{equation}\label{eq:betaphi}
\beta\left(\phi\right)\,=\,\frac{\dd \phi}{\dd A}\,=\,-2\left(d-1\right)\frac{\dd\log W}{\dd\phi}\,.
\end{equation}
Close to any of the three real FPs $\phi_c=\{ 0,\phi_0,\overline{\phi}_0\}$ one finds the expected behaviour 
\be\label{UVRG}
\phi \simeq g_c \,e^{-(d-\Delta_c)\rho/L_c}
\,, \qquad A(\rho)\simeq \frac{\rho}{L_c}\sim \log \frac{\mu}{\Lambda_c},
\ee
and 
\be
\beta(\phi)\simeq -(d-\Delta_c)(\phi-\phi_c)+ O \left((\phi-\phi_c)^2\right)\,,
\ee
where $\mu$ is the RG scale, $\Lambda_c$ is the scale that triggers the flow away or into the FP, and $g_c$ is the corresponding (dimensionless) coupling at the FP. UV and IR FPs are approached for $\rho \to \infty$ and $\rho \to -\infty$, respectively. The $\beta$-functions for our model are shown in Fig.~\ref{fig:beta}, where we see that they exhibit the  behaviour discussed in Sec.~\ref{sec:FPAandCCFT}.

Since the scalar field is dual to the coupling constant in the gauge theory, we propose that the holographic dual of the extension of this coupling to complex values consists of extending the scalar field on the gravity side to complex values as well. Since the scalar field couples to the metric, we also extend the metric components to complex values. We assume that, in this extension, the action \eqref{eq:action} is a holomorphic function of $g_{MN}$ and $\phi$. In other words, we do not introduce any explicit dependence on the complex conjugates of these fields in the action. Put yet another way, the equations of motion are obtained by varying the action with respect to $\phi$ and $g_{MN}$ as complex variables, as opposed to varying independently with respect to their real and imaginary parts.

With this extension, the FPs do not disappear at $\overline{\phi}_0=\phi_0$ but simply move to the complex-$\phi$ plane. The dimensions of the operators dual to the scalar field at each CFP \eqref{eq:D0W} are complex conjugate of one another, $\overline{\Delta}_0=\Delta_0^*$. In addition, there are formally AdS solutions with metrics 
$g_{MN}=\{ h_{MN},\overline{h}_{MN} \}$ whose complex radii \eqref{eq:L0W} are also complex conjugates, $\overline{L}_0=L_0^*$. Since we assume that the coordinates are real, this relates the two metrics by complex conjugation $\overline{h}_{MN}=h_{MN}^*$. Accordingly, any quantities that can be computed holographically at the CFPs purely in terms of geometric quantities will be related by complex conjugation. These include the expectation value of Wilson loops \cite{Maldacena:1998im,Rey:1998bq}, the entanglement entropy \cite{Ryu:2006bv},  and holographic $c$-functions \cite{Freedman:1999gp} that are related to central charges and anomaly coefficients.

\begin{figure}[t]
	\begin{center}
			\includegraphics[width=0.45\textwidth,height=0.3\textwidth]{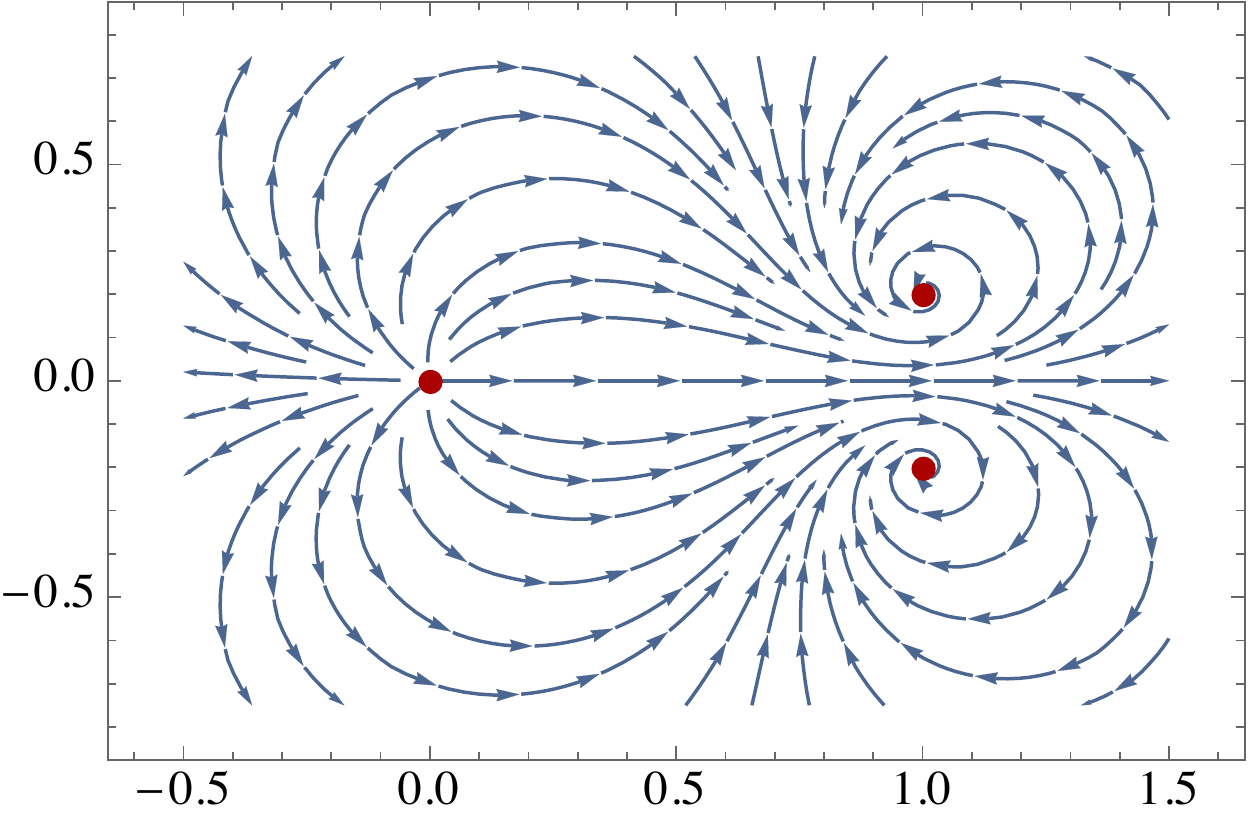} 
			\put(-180,123){$\operatorname{Im}\phi $}
			\put(-22,15){$ \operatorname{Re}\phi$}
		\caption{\small  Examples of complex RG flows for $d=4$ displaying the characteristic spiralling behavior around the CFPs. The flow in the UV is triggered by a source for a $\Delta_{\text{\tiny UV}}=3$ operator and the CFPs are located at $\phi_0=\overline{\phi}_0^*=1+0.2i$.}\label{fig:RGflow}
	\end{center}
\end{figure}

This result extends to complex RG flows between the UV FP at $\phi=0$ and the CFPs. Once we continue the scalar to complex values, the first-order equations \eqref{BPS} split into real and imaginary parts and the solutions describe the RG flow of the real and imaginary parts of the dual coupling. The scalar field still approaches  the CFPs as given by the first equation in \eqref{UVRG}. Since both $\Delta_c$ and $L_c$ are complex the coupling oscillates. In the particular example of Fig.~\ref{fig:RGflow} the CFPs are IR FPs but this is not generic, i.e.~CFPs can also be UV FPs. The only purely real flow is the straight horizontal line that passes exactly in between the CFPs and should exhibit walking behavior.  We have collected some explicit formulas for the flows in Appendix \ref{sec:complexRGflows}.

\subsection{Holographic complex conformal field theories}
\label{sec:holoCCFT}

The holographic CFPs show many of the properties expected for a pair of CCFTs. In addition to the operator dual to $\phi$, we will show that the spectrum of dimensions of local operators at one CFP is the complex conjugate of its companion's. Let $X^I$ denote the real components of a field in an arbitrary Lorentz representation. Then the action expanded to quadratic order around a CFP at $\phi_c=\{\phi_0,\overline{\phi}_0\}$, $h_c=\{ h,\overline{h}\}$ will be
\be
{\mathcal L}_c\simeq - \sqrt{-h_c}\left( \frac{1}{2}X^I K_{c\,IJ} X^J+\frac{1}{2}M^2_{c\,IJ} X^I X^J\right),
\ee
where we have separated a kinetic part determined by a differential operator $K_{c\,IJ}=K_{IJ}(h_c,\phi_c)=\{ K_{IJ},\overline{K}_{IJ}\}$ and a mass term $M^2_{c\,IJ}=M^2_{IJ}(h_c, \phi_c)=\{ M^2_{IJ},\overline{M}^2_{IJ}\}$. This gives the field equations
\be\label{eq:eom}
K_{c\,IJ} X_c^J+M^2_{c\,IJ} X_c^J=0.
\ee
The general solution near the AdS boundary  will be a superposition of exponentials of the form \footnote{For integer valued $\Delta_{c\,n}$ there can be additional powers of $\rho$ in the solutions, but the exponents do not change.}
\be\label{eq:XI}
X_c^I=\sum_n a_{c\,n}^I e^{-\Delta_{c\,n} \rho/L_c}+b_{c\,n}^Ie^{-(d-\Delta_{c\,n}) \rho/L_c}.
\ee
Introducing this in \eqref{eq:eom} one finds a homogeneous system of equations for the coefficients $a_{c\,n}^I$, $b_{c\,n}^I$ that has solutions when $\Delta_{c\,n}=\{ \Delta_n,\overline{\Delta}_n\}$ take the values corresponding to the conformal dimensions of the dual operators $\cO_{c\,n}=\{ \cO_n,\overline{\cO}_n\}$. 
We can now use holomorphicity of the action to show that 
\be
\overline{K}=K(\overline{h},\overline{\phi}_0)=K(h^*,\phi_0^*)=  \left( K(h,\phi_0) \right)^*=(K)^*
\ee
and, similarly, that  $\overline{M}^2=(M^2)^*$, where we have suppressed the $IJ$-indices for simplicity.
This implies that the spectra of operators at the two CFPs are related by complex conjugation, as anticipated:
\be
\overline{\Delta}_n=\Delta_n^*.
\ee

\subsection{Walking behavior and Miransky scaling}
\label{sec:walking}

Our simple holographic model correctly describes the physics of walking and the associated Miransky scaling when the CFPs are close to the real axis. In particular, the $\beta$-function \eqref{eq:betaphi} reproduces  \eqref{eq:BKT}. We will illustrate this scaling further by heating up the real RG flow that passes exactly in between the CFPs in Fig.~\ref{fig:RGflow}. On the gravity side this corresponds to constructing black hole solutions that start in the UV as deformations of a $d=4$ CFT by a $\Delta_{\text{\tiny UV}}=3$ operator with source $\Lambda$. We set $\phi_0= 1+i\epsilon$ and construct  solutions for several small values of $\epsilon$, following the procedure described in \cite{Gubser:2008ny}. Some explicit technical details can be found in Appendix \ref{sec:BH}.

\begin{figure}[t]
	\begin{center}
			\includegraphics[width=0.45\textwidth,height=0.3\textwidth]{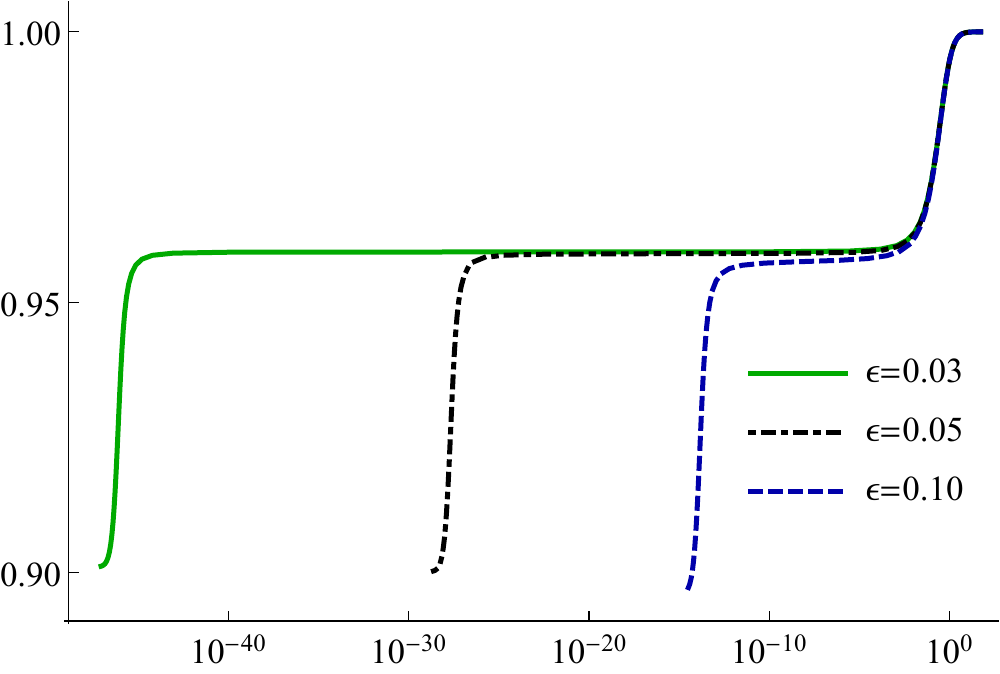} 
				\put(-180,125){$ S/S_{\text{\tiny UV}}$}
			\put(-17,17){$ T/\Lambda$}
		\caption{\small Log-linear plots of the entropy density $S$ as a function of the temperature $T$, normalized to that of the UV CFT, for different values of $\phi_0= 1+ i\epsilon$.}\label{fig:entropy}
	\end{center}
\end{figure}

\begin{figure}[t]
	\begin{center}
			\includegraphics[width=0.45\textwidth,height=0.3\textwidth]{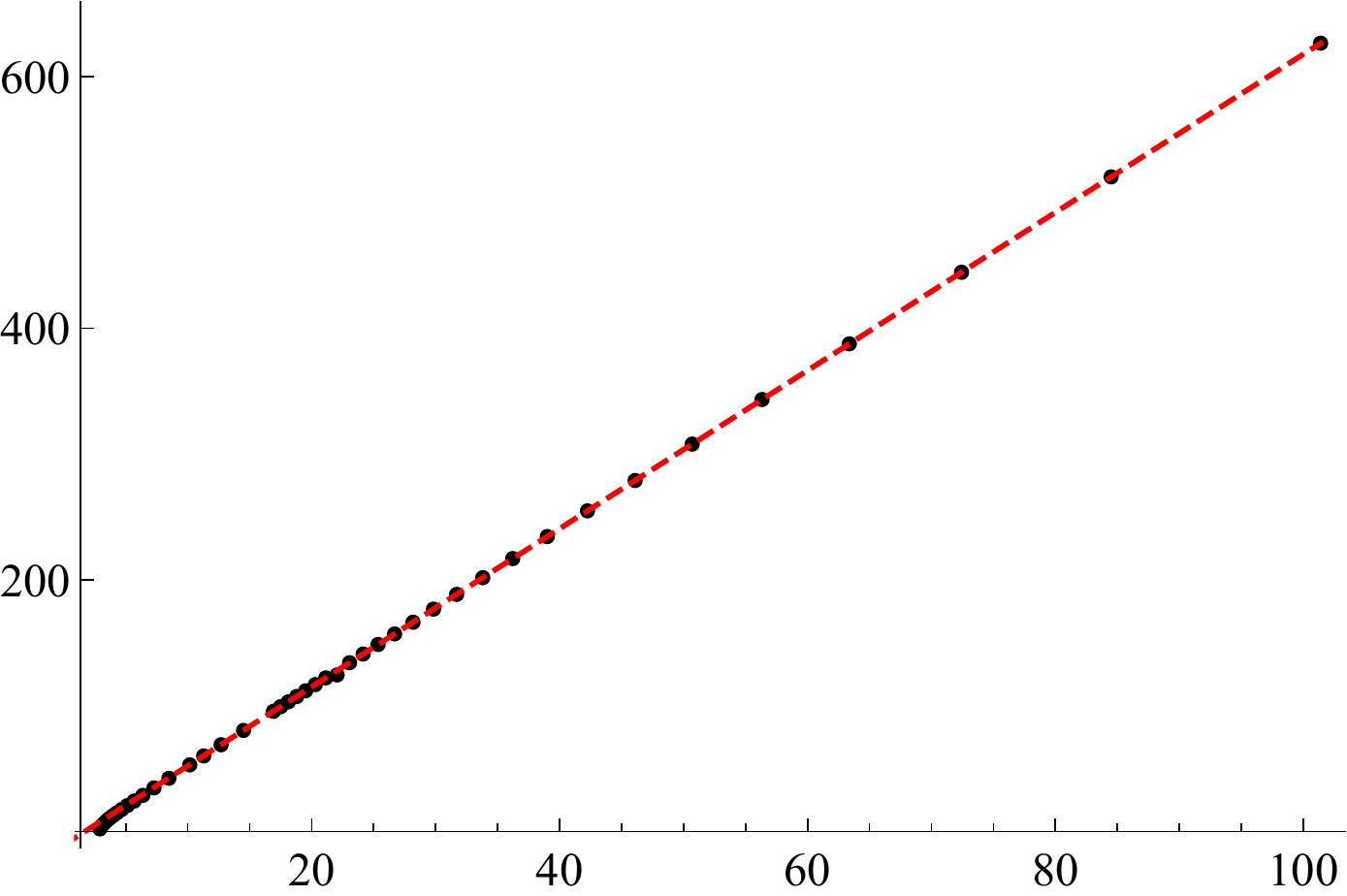} 
			\put(-182,125){$\log\frac{\mu_{\text{\tiny UV}}}{\mu_{\text{\tiny IR}}} $}
			\put(-40,18){$ |\operatorname{Im}\Delta_0|^{-1}$}
		\caption{\small  Size of the walking region as a function of the imaginary part of the dimension of the operator at the CFP. The black dots are the values computed in our model while the red dashed line  is a line with slope $2\pi$ that passes through the last black dot.}\label{fig:Miransky}
	\end{center}
\end{figure}

At very high temperature the thermodynamics is dominated by the physics of the UV CFT, hence $S_{\text{\tiny UV}} \propto T^3$. In a region with walking the entropy would show a similar temperature scaling, so $S/S_{\text{\tiny UV}}$ should be approximately constant. This plateaux can be clearly seen in the log-linear plot of  Fig.~\ref{fig:entropy}. 
To quantify the size of the plateaux, we declare that the flow is in the walking region if the following derivative is smaller than a certain control parameter $\nu$
\begin{equation}
\frac{\dd \log(S/S_{UV})}{\dd \log (T/\Lambda)} < \nu\,. 
\end{equation}
For definiteness we take $\nu=\frac{1}{2}10^{-3}$. The condition is satisfied for values of the temperature in a bounded range $T_{\text{\tiny UV}}>T>T_{\text{\tiny IR}}$.  We identify these temperatures with the energy scales $\mu_{\text{\tiny UV}}$ and $\mu_{\text{\tiny IR}}$. We then vary 
$\epsilon$ and plot the ratio of these scales as a function of the imaginary part of the scaling dimension $\Delta_0$ at the CFP. The result is 
Fig.~\ref{fig:Miransky}, which exhibits the expected scaling \eqref{Miransky}.

\section{Discussion}

The simple holographic model that we have presented captures the physics of FPA. Continuing the scalar field to complex values in such a way that the action remains holomorphic, it is also possible to describe CFPs. Then a straightforward extension of the rules of the gauge/gravity duality allows us to study not only the properties of conjectured CCFTs at strong coupling but also the complex RG flows between them. When the CFP are close to the real axis the real RG flow that passes exactly between them walks and displays the associated Miransky scaling behavior. 

The holomorphic gravitational action defined in this way is complex and non-Hermitian, so its meaning beyond the classical level is  unclear. It would be nice to relate our proposal to those in \cite{Witten:2010zr,Behtash:2015zha}, where it was argued that complex saddle points may give important contributions to the path integral in particular cases. 

Our work shows that, contrary to some belief in the community, the RG flow leading to FPA in large-$N_c$ theories may be driven by a single-trace operator, the operator dual to the scalar field. Our model avoids the argument in Appendix D of \cite{Gorbenko:2018ncu} because the the  coupling constant associated to this operator is O($N_c$) instead of O(1).

Our construction  is based on a bottom-up model. It would  be interesting to find top-down, string theory realizations of FPA and CFPs. While the scalar potentials obtained in consistent truncations of string and M-theory generically possess CFPs, an open question is whether their distance to the real axis is controlled by some parameter that can be freely varied. This issue is currently under investigation.


\section*{Acknowledgements}
We are grateful to Y.~Bea for collaboration in the initial stages of this project. We thank B.~Fiol, I.~Herbut, A.~Pomarol and O.~Pujolas for discussions. 
AF and DM are supported by grants FPA2016-76005-C2-1-P, FPA2016-76005-C2-2-P, SGR-2017-754 and MDM-2014-0369. C.H. is partially supported by the grants PGC2018-096894-B-100, GRUPIN-18-174 and the Ramon y Cajal fellowship RYC-2012-10370. JGS acknowledges support from the FPU program, fellowship FPU15/02551.

\appendix
\setcounter{tocdepth}{1}

\section{Complex RG flows}
\label{sec:complexRGflows}

The $\beta$-function \eqref{eq:betaphi} corresponding to \eqref{dW} has the desired properties to describe both walking and the physics of cCFTs. Expanding to quadratic order in $g\equiv\phi-\phi_0$ around the fixed point at $\phi=\phi_0$ we find
\begin{equation}\label{betaW}
\beta\,=\,-2\left(d-1\right)W_0\phi_0\left(\phi_0-\overline{\phi}_0\right)\frac{L_0}{L}\,g-2\left(d-1\right)W_0\phi_0\left(2\phi_0-\overline{\phi}_0\right)\frac{L_0}{L}\,g^2+O(g^3)\,,
\end{equation}
where $L_0$ is given in \eqref{eq:L0W}. To this quadratic order, in addition to the fixed point at $g=0$, there is another one at
\begin{equation}
\overline{g}\,\equiv\,-\frac{\phi_0\left(\phi_0-\overline{\phi}_0\right)}{2\phi_0-\overline{\phi}_0}\,.
\end{equation}
When the fixed point at $\phi=\phi_0$ is close to the real axis, i.e.~if 
$\phi_0=1+i \,\epsilon$ with $\epsilon\ll1$, then this second fixed point is its complex conjugate since
\be
\overline{\phi}_0 = \phi_0 + \overline{g} = 1-i\epsilon + O(\epsilon^2)
\ee
Under these circumstances  the model is expected to exhibit walking behaviour when the flow passes exactly between the two complex fixed points. In this regime the $\beta$-function is obtained by evaluating \eqref{betaW} at $g_{\text{\tiny W}}+\overline{g}/2$ with $g_{\text{\tiny W}}$ real. By rescaling the coupling $g_{\text{\tiny W}}$, it can be seen that this $\beta$-function is of the form \eqref{eq:beta} with
\begin{equation}
\alpha-\alpha_*\,=\,-\frac{576\left(d-1\right)^2W_0^2}{\left(12+W_0\right)^2}\,\epsilon^2\,=\,-\frac14|\operatorname{Im}{\Delta_0}|^2\,,
\end{equation}
recovering the results of the previous sections.

It is also straightforward to study the complex flows of this model by assuming that $\phi$ is a complex field so that the action is a holomorphic function. Both the field and the complex extrema of the potential have now both real and imaginary parts
\begin{equation}
\phi\,=\,\phi_{\tiny{\rm R}}+i\,\phi_{\tiny{\rm I}}\,,\qquad\qquad\qquad\phi_0\,=\,\phi_0^{\tiny{\rm R}}+i\,\phi_0^{\tiny{\rm I}}\,.
\end{equation}
In terms of these the  first-order equations \eqref{BPS} take the form
\begin{equation}\label{complexflows}
\begin{array}{rcl}
\phi_{\tiny{\rm R}}'&=&\frac{1}{L|\phi_0|^2}\left[\left(3\phi_{\tiny{\rm I}}^2+2\phi_0^{\tiny{\rm R}}\phi_{\tiny{\rm R}}-|\phi_0|^2-\phi_{\tiny{\rm R}}^2\right)\phi_{\tiny{\rm R}}-2\phi_{\tiny{\rm I}}^2\phi_0^{\tiny{\rm R}}\right]\,,\\[4mm]
\phi_{\tiny{\rm I}}'&=&\frac{1}{L|\phi_0|^2}\left(\phi_{\tiny{\rm I}}^2-|\phi_0|^2-3\phi_{\tiny{\rm R}}^2+4\phi_0^{\tiny{\rm R}}\phi_{\tiny{\rm R}}\right)\phi_{\tiny{\rm I}}\,,
\end{array}
\end{equation}
where we fixed $W_0$ so that the scalar is dual to an operator of dimension $\Delta_\text{\tiny UV}=3$ at the UV fixed point and moreover we are working in a four-dimensional gauge theory.
At the linear level around the UV fixed point, the system is solved by
\begin{equation}
\delta\phi_{\tiny{\rm R}}\,=\,v_{\tiny{\rm R}}\,e^{-\frac{\rho}{L}}\,,\qquad\qquad\delta\phi_{\tiny{\rm I}}\,=\,v_{\tiny{\rm I}}\,e^{-\frac{\rho}{L}}\,,
\end{equation}
showing that both the real and imaginary parts correspond indeed to sources for a $\Delta_\text{\tiny UV}=3$ operator. On the other hand, around $\phi=\phi_0$ the linear solution is
\begin{equation}
\begin{array}{rcl}
\delta\phi_{\tiny{\rm R}}&=&c_1\,e^{\frac{2(\phi_0^{\tiny{\rm I}})^2}{|\phi_0|^2}\frac{\rho}{L}} \cos\left(\frac{2\phi_0^{\tiny{\rm R}}\phi_0^{\tiny{\rm I}}}{|\phi_0|^2}\frac{\rho}{L}\right) +c_2\,e^{\frac{2(\phi_0^{\tiny{\rm I}})^2}{|\phi_0|^2}\frac{\rho}{L}} \sin\left(\frac{2\phi_0^{\tiny{\rm R}}\phi_0^{\tiny{\rm I}}}{|\phi_0|^2}\frac{\rho}{L}\right)  \,,\\[2mm]
\delta\phi_{\tiny{\rm I}}&=&c_2\,e^{\frac{2(\phi_0^{\tiny{\rm I}})^2}{|\phi_0|^2}\frac{\rho}{L}} \cos\left(\frac{2\phi_0^{\tiny{\rm R}}\phi_0^{\tiny{\rm I}}}{|\phi_0|^2}\frac{\rho}{L}\right) -c_1\,e^{\frac{2(\phi_0^{\tiny{\rm I}})^2}{|\phi_0|^2}\frac{\rho}{L}} \sin\left(\frac{2\phi_0^{\tiny{\rm R}}\phi_0^{\tiny{\rm I}}}{|\phi_0|^2}\frac{\rho}{L}\right) \,,
\end{array}
\end{equation}
for real constants $c_1$ and $c_2$. Notice that the modulus is exponentially decreasing as $\rho\to-\infty$, while the phase has an infinite number of oscillations. Examples of full non-linear flows are depicted in Figure \ref{fig:RGflow}. The result is similar to the cartoon in Figure 11 of \cite{Gorbenko:2018dtm}.

\section{Black hole solutions}
\label{sec:BH}

In order to construct the black hole solutions we followed the procedure described in \cite{Gubser:2008ny}, which we now review for any dimensionality. This is applicable to any gravitational action of the form \eqref{eq:action} with arbitrary potential admitting a UV fixed point. 

We are interested in solutions which are asymptotically AdS$_{d+1}$, so we choose the following ansatz for the metric
\begin{equation}\label{eq:Ansatz}
\dd s^2_{d+1} = e^{2A} \parent{- h\  \dd t^2 + \dd \vec{x}^2} + e^{2B} \ \frac{\dd r^2}{h}\,,
\end{equation}
where $\dd\vec{x}^2 = \dd x_1^2+\cdots + \dd x_{d-1}^2$ and we are assuming that all the functions depend only on the radial coordinate. Note that, since we are including the $e^{2B}$ factor, the radial gauge is not fixed yet. It is possible to use this freedom to consider the scalar $\phi$ as the radial coordinate, so that
\begin{equation}
\label{eq:ansatzPhi}
\dd s^2_{d+1} = e^{2A} \parent{- h\  \dd t^2 + \dd \vec{x}^2} + e^{2B} \ \frac{\dd \phi^2}{h}\,.
\end{equation}
The price to pay is a dynamical equation determining $B$. With this choice it is implicitly assumed that $\phi$ is either monotonically increasing or decreasing. Then, from \eqref{eq:action} and \eqref{eq:ansatzPhi} we obtain the following equations of motion
\begin{equation}\label{eq:EOM}
\begin{aligned}
0&=h'' + \parent{d A' - B'}h'\,,\\[2mm]
0&=A'' -A'B' + \frac{1}{2(d-1)}\,,\\[2mm]
0&=d\ A'-B' +\frac{h'}{h} -\frac{e^{2B}}{h}V'\,,\\[2mm]
0&=2(d-1)\ A'h'+h\ \parent{\ 2d (d-1)\ A'^2 -1}+2e^{2B} V\,,
\end{aligned}
\end{equation}
where primes denote differentiation with respect to $\phi$. It turns out to be useful to define the function $G(\phi) = A'(\phi)$, in such a way that the solution can be given in terms of the following integrals involving $G$
\begin{equation}
\label{eq:solution}\begin{aligned}
A(\phi)&= A_0+\int_{\phi_0}^{\phi} \dd\tilde{\phi} \ G(\tilde{\phi})\,,\\
B(\phi)&= B_0+\int_{\phi_0}^{\phi} \dd\tilde{\phi}\  \frac{1}{G(\tilde{\phi})} \parent{ {G'(\tilde{\phi})+\frac{1}{2(d-1)}}}\,,\\
h(\phi)&= h_0 + h_1\int_{\phi_0}^{\phi}\ \dd\tilde{\phi}\ e^{-d A(\tilde{\phi}) + B(\tilde{\phi})}\,.
\end{aligned}
\end{equation}
Moreover, there is a relation between the potential and this new function 
\begin{equation}\label{eq:potential}
V(\phi) = \frac{h}{2} e^{-2B}\ \parent{1-2d(d-1)\  G^2 -2(d-1) \ G\ \frac{h'}{h}}\,.
\end{equation}
By differentiating some combinations of \eqref{eq:solution} and \eqref{eq:potential} it can be seen that $G$ satisfies the following master equation 
\begin{equation}
\label{ap:eq:equationOfG}
\frac{G'(\phi)}{G(\phi)+ \frac{V(\phi)}{(d-1)\ V'(\phi)}} \,=\, \frac{\dd}{\dd \phi}\left[\frac{G'(\phi)}{G(\phi)} + \frac{1}{2(d-1)G(\phi)}-d\  G(\phi) - \frac{G'(\phi)}{G(\phi)+ \frac{V(\phi)}{(d-1)\ V'(\phi)}}\right]\,.
\end{equation} 
As usual, the boundary conditions at the horizon (placed at $\phi=\phi_H$) are such that $h$ has a simple zero, whereas the remaining functions are finite. In order to see what conditions should be imposed on $G$ at the horizon, we evaluate the last two equations in \eqref{eq:EOM} to obtain 
\begin{equation}
V(\phi_H)\,=\,-(d-1)\  e^{-2B(\phi_H)}\ G(\phi_H)h'(\phi_H)\,,\qquad\qquad V'(\phi_H) \,=\, e^{-2B(\phi_H)}h'(\phi_H)\,.
\end{equation}
Consequently, $G+\frac{V}{(d-1)V'}$ must vanish at the horizon. Taking this into account, \eqref{ap:eq:equationOfG} can be solved perturbatively near the horizon
\begin{equation}
\label{eq:GnearH}
G(\phi)\,=\, -\frac{1}{d-1}\frac{V(\phi)}{V'(\phi)}+ \frac{1}{2(d-1)}\parent{\frac{V(\phi_H)V''(\phi_H)}{V'(\phi_H)^2}-1}(\phi-\phi_H) + O\left[(\phi-\phi_H)^2\right] \,.
\end{equation}
This series can be extended to any order without finding further integration constants. The numerical strategy to solve \eqref{ap:eq:equationOfG} is as follows. We choose a value $\phi_H$ for the scalar and evaluate \eqref{eq:GnearH} near the horizon. Then this value is introduced as a seed in a numerical integrator such as Mathematica's \verb|NDSolve|, which can be used to solve \eqref{ap:eq:equationOfG} numerically up to a value close to the UV fixed point, corresponding to AdS.

The temperature $T$ and entropy $S$ associated to these black holes are computed as usual from the Euclidean time period and the horizon area, respectively. Indeed, they can be written as simple integrals involving $G$
\begin{equation}
\label{eq:SandT}
\begin{aligned}
T&= \frac{d\ \phi_H^{1/(\Delta-d)}}{4\pi L}  \frac{V(\phi_H)}{V(0)}\exp \left\{ \int_0^{\phi_H}\dd \phi \left[G(\phi)-\frac{1}{(\Delta-d)\phi}+\frac{1}{2(d-1)G(\phi)}\right]\right\}\,,\\[2mm]
S&= \frac{2\pi}{2\kappa_{d+1}^2} \phi_H^{\frac{d-1}{\Delta-d}} \exp\left\{ (d-1)\int_0^{\phi_H} \dd\phi \ \left[G(\phi) - \frac{1}{(\Delta-d)\phi}\right]  \right\}\,,
\end{aligned}
\end{equation}
where the appropriate values for the integration constants appearing in \eqref{eq:solution} were chosen so that the same AdS asymptotics is recovered for the different $\phi_H$ that we consider (see \cite{Gubser:2008ny} for details).

\bibliographystyle{JHEP}
\bibliography{refscCFT}

\providecommand{\href}[2]{#2}\begingroup\raggedright\begin{thebibliography}{10}

\bibitem{Halperin:1973jh}
B.~I. Halperin, T.~C. Lubensky and S.-k. Ma, \emph{{First order phase
  transitions in superconductors and smectic A liquid crystals}},
  \href{https://doi.org/10.1103/PhysRevLett.32.292}{\emph{Phys. Rev. Lett.}
  {\bfseries 32} (1974) 292}.

\bibitem{Ihrig:2019kfv}
B.~Ihrig, N.~Zerf, P.~Marquard, I.~F. Herbut and M.~M. Scherer, \emph{{Abelian
  Higgs model at four loops, fixed-point collision and deconfined
  criticality}}, \href{https://doi.org/10.1103/PhysRevB.100.134507}{\emph{Phys.
  Rev.} {\bfseries B100} (2019) 134507}
  [\href{https://arxiv.org/abs/1907.08140}{{\ttfamily 1907.08140}}].

\bibitem{Senthil:2004}
T.~{Senthil}, A.~{Vishwanath}, L.~{Balents}, S.~{Sachdev} and M.~P.~A.
  {Fisher}, \emph{{Deconfined Quantum Critical Points}},
  \href{https://doi.org/10.1126/science.1091806}{\emph{Science} {\bfseries 303}
  (2004) 1490} [\href{https://arxiv.org/abs/cond-mat/0311326}{{\ttfamily
  cond-mat/0311326}}].

\bibitem{Nahum:2015jya}
A.~Nahum, J.~T. Chalker, P.~Serna, M.~Ortu\~no and A.~M. Somoza,
  \emph{{Deconfined Quantum Criticality, Scaling Violations, and Classical Loop
  Models}}, \href{https://doi.org/10.1103/PhysRevX.5.041048}{\emph{Phys. Rev.}
  {\bfseries X5} (2015) 041048}
  [\href{https://arxiv.org/abs/1506.06798}{{\ttfamily 1506.06798}}].

\bibitem{Wang:2017txt}
C.~Wang, A.~Nahum, M.~A. Metlitski, C.~Xu and T.~Senthil, \emph{{Deconfined
  quantum critical points: symmetries and dualities}},
  \href{https://doi.org/10.1103/PhysRevX.7.031051}{\emph{Phys. Rev.} {\bfseries
  X7} (2017) 031051} [\href{https://arxiv.org/abs/1703.02426}{{\ttfamily
  1703.02426}}].

\bibitem{Serna:2018tct}
P.~Serna and A.~Nahum, \emph{{Emergence and spontaneous breaking of approximate
  $O(4)$ symmetry at a weakly first-order deconfined phase transition}},
  \href{https://doi.org/10.1103/PhysRevB.99.195110}{\emph{Phys. Rev.}
  {\bfseries B99} (2019) 195110}
  [\href{https://arxiv.org/abs/1805.03759}{{\ttfamily 1805.03759}}].

\bibitem{Nienhuis:1979mb}
B.~Nienhuis, A.~N. Berker, E.~K. Riedel and M.~Schick, \emph{{First and Second
  Order Phase Transitions in Potts Models: Renormalization - Group Solution}},
  \href{https://doi.org/10.1103/PhysRevLett.43.737}{\emph{Phys. Rev. Lett.}
  {\bfseries 43} (1979) 737}.

\bibitem{Nauenberg:1980nv}
M.~Nauenberg and D.~J. Scalapino, \emph{{Singularities and Scaling Functions at
  the Potts Model Multicritical Point}},
  \href{https://doi.org/10.1103/PhysRevLett.44.837}{\emph{Phys. Rev. Lett.}
  {\bfseries 44} (1980) 837}.

\bibitem{Gorbenko:2018dtm}
V.~Gorbenko, S.~Rychkov and B.~Zan, \emph{{Walking, Weak first-order
  transitions, and Complex CFTs II. Two-dimensional Potts model at $Q>4$}},
  \href{https://doi.org/10.21468/SciPostPhys.5.5.050}{\emph{SciPost Phys.}
  {\bfseries 5} (2018) 050} [\href{https://arxiv.org/abs/1808.04380}{{\ttfamily
  1808.04380}}].

\bibitem{Herbut:2014lfa}
I.~F. Herbut and L.~Janssen, \emph{{Topological Mott insulator in
  three-dimensional systems with quadratic band touching}},
  \href{https://doi.org/10.1103/PhysRevLett.113.106401}{\emph{Phys. Rev. Lett.}
  {\bfseries 113} (2014) 106401}
  [\href{https://arxiv.org/abs/1404.5721}{{\ttfamily 1404.5721}}].

\bibitem{Fei:2014xta}
L.~Fei, S.~Giombi, I.~R. Klebanov and G.~Tarnopolsky, \emph{{Three loop
  analysis of the critical O(N) models in 6-$\epsilon$ dimensions}},
  \href{https://doi.org/10.1103/PhysRevD.91.045011}{\emph{Phys. Rev.}
  {\bfseries D91} (2015) 045011}
  [\href{https://arxiv.org/abs/1411.1099}{{\ttfamily 1411.1099}}].

\bibitem{Gracey:2018khg}
J.~A. Gracey, I.~F. Herbut and D.~Roscher, \emph{{Tensor $O(N)$ model near six
  dimensions: fixed points and conformal windows from four loops}},
  \href{https://doi.org/10.1103/PhysRevD.98.096014}{\emph{Phys. Rev.}
  {\bfseries D98} (2018) 096014}
  [\href{https://arxiv.org/abs/1810.05721}{{\ttfamily 1810.05721}}].

\bibitem{Appelquist:1988sr}
T.~Appelquist, D.~Nash and L.~C.~R. Wijewardhana, \emph{{Critical Behavior in
  (2+1)-Dimensional QED}},
  \href{https://doi.org/10.1103/PhysRevLett.60.2575}{\emph{Phys. Rev. Lett.}
  {\bfseries 60} (1988) 2575}.

\bibitem{Kubota:2001kk}
K.-i. Kubota and H.~Terao, \emph{{Dynamical symmetry breaking in QED(3) from
  the Wilson RG point of view}},
  \href{https://doi.org/10.1143/PTP.105.809}{\emph{Prog. Theor. Phys.}
  {\bfseries 105} (2001) 809}
  [\href{https://arxiv.org/abs/hep-ph/0101073}{{\ttfamily hep-ph/0101073}}].

\bibitem{Kaveh:2004qa}
K.~Kaveh and I.~F. Herbut, \emph{{Chiral symmetry breaking in QED(3) in
  presence of irrelevant interactions: A Renormalization group study}},
  \href{https://doi.org/10.1103/PhysRevB.71.184519}{\emph{Phys. Rev.}
  {\bfseries B71} (2005) 184519}
  [\href{https://arxiv.org/abs/cond-mat/0411594}{{\ttfamily
  cond-mat/0411594}}].

\bibitem{Herbut:2016ide}
I.~F. Herbut, \emph{{Chiral symmetry breaking in three-dimensional quantum
  electrodynamics as fixed point annihilation}},
  \href{https://doi.org/10.1103/PhysRevD.94.025036}{\emph{Phys. Rev.}
  {\bfseries D94} (2016) 025036}
  [\href{https://arxiv.org/abs/1605.09482}{{\ttfamily 1605.09482}}].

\bibitem{Gies:2005as}
H.~Gies and J.~Jaeckel, \emph{{Chiral phase structure of QCD with many
  flavors}}, \href{https://doi.org/10.1140/epjc/s2006-02475-0}{\emph{Eur. Phys.
  J.} {\bfseries C46} (2006) 433}
  [\href{https://arxiv.org/abs/hep-ph/0507171}{{\ttfamily hep-ph/0507171}}].

\bibitem{Pomoni:2008de}
E.~Pomoni and L.~Rastelli, \emph{{Large N Field Theory and AdS Tachyons}},
  \href{https://doi.org/10.1088/1126-6708/2009/04/020}{\emph{JHEP} {\bfseries
  04} (2009) 020} [\href{https://arxiv.org/abs/0805.2261}{{\ttfamily
  0805.2261}}].

\bibitem{Kaplan:2009kr}
D.~B. Kaplan, J.-W. Lee, D.~T. Son and M.~A. Stephanov, \emph{{Conformality
  Lost}}, \href{https://doi.org/10.1103/PhysRevD.80.125005}{\emph{Phys. Rev.}
  {\bfseries D80} (2009) 125005}
  [\href{https://arxiv.org/abs/0905.4752}{{\ttfamily 0905.4752}}].

\bibitem{Antipin:2012kc}
O.~Antipin, S.~Di~Chiara, M.~Mojaza, E.~M{\o}lgaard and F.~Sannino, \emph{{A
  Perturbative Realization of Miransky Scaling}},
  \href{https://doi.org/10.1103/PhysRevD.86.085009}{\emph{Phys. Rev.}
  {\bfseries D86} (2012) 085009}
  [\href{https://arxiv.org/abs/1205.6157}{{\ttfamily 1205.6157}}].

\bibitem{Hansen:2017pwe}
F.~F. Hansen, T.~Janowski, K.~Lang{\ae}ble, R.~B. Mann, F.~Sannino, T.~G.
  Steele et~al., \emph{{Phase structure of complete asymptotically free
  SU($N_c$) theories with quarks and scalar quarks}},
  \href{https://doi.org/10.1103/PhysRevD.97.065014}{\emph{Phys. Rev.}
  {\bfseries D97} (2018) 065014}
  [\href{https://arxiv.org/abs/1706.06402}{{\ttfamily 1706.06402}}].

\bibitem{Gorbenko:2018ncu}
V.~Gorbenko, S.~Rychkov and B.~Zan, \emph{{Walking, Weak first-order
  transitions, and Complex CFTs}},
  \href{https://doi.org/10.1007/JHEP10(2018)108}{\emph{JHEP} {\bfseries 10}
  (2018) 108} [\href{https://arxiv.org/abs/1807.11512}{{\ttfamily
  1807.11512}}].

\bibitem{Miransky:1984ef}
V.~A. Miransky, \emph{{Dynamics of Spontaneous Chiral Symmetry Breaking and
  Continuum Limit in Quantum Electrodynamics}},
  \href{https://doi.org/10.1007/BF02724229}{\emph{Nuovo Cim.} {\bfseries A90}
  (1985) 149}.

\bibitem{Berezinsky:1970fr}
V.~L. Berezinsky, \emph{{Destruction of long range order in one-dimensional and
  two-dimensional systems having a continuous symmetry group. I. Classical
  systems}}, {\emph{Sov. Phys. JETP} {\bfseries 32} (1971) 493}.

\bibitem{Kosterlitz:1973xp}
J.~M. Kosterlitz and D.~J. Thouless, \emph{{Ordering, metastability and phase
  transitions in two-dimensional systems}},
  \href{https://doi.org/10.1088/0022-3719/6/7/010}{\emph{J. Phys.} {\bfseries
  C6} (1973) 1181}.

\bibitem{Sieg:2016vap}
C.~Sieg and M.~Wilhelm, \emph{{On a CFT limit of planar $\gamma_i$-deformed
  $\mathcal{N}=4$ SYM theory}},
  \href{https://doi.org/10.1016/j.physletb.2016.03.004}{\emph{Phys. Lett.}
  {\bfseries B756} (2016) 118}
  [\href{https://arxiv.org/abs/1602.05817}{{\ttfamily 1602.05817}}].

\bibitem{Grabner:2017pgm}
D.~Grabner, N.~Gromov, V.~Kazakov and G.~Korchemsky, \emph{{Strongly
  $\gamma$-Deformed $\mathcal{N}=4$ Supersymmetric Yang-Mills Theory as an
  Integrable Conformal Field Theory}},
  \href{https://doi.org/10.1103/PhysRevLett.120.111601}{\emph{Phys. Rev. Lett.}
  {\bfseries 120} (2018) 111601}
  [\href{https://arxiv.org/abs/1711.04786}{{\ttfamily 1711.04786}}].

\bibitem{Pittelli:2019ceq}
A.~Pittelli and M.~Preti, \emph{{Integrable Fishnet from $\gamma$-Deformed
  $\mathcal{N}=2$ Quivers}},
  \href{https://doi.org/10.1016/j.physletb.2019.134971}{\emph{Phys. Lett.}
  {\bfseries B798} (2019) 134971}
  [\href{https://arxiv.org/abs/1906.03680}{{\ttfamily 1906.03680}}].

\bibitem{Benini:2019dfy}
F.~Benini, C.~Iossa and M.~Serone, \emph{{Conformality Loss, Walking, and 4D
  Complex Conformal Field Theories at Weak Coupling}},
  \href{https://doi.org/10.1103/PhysRevLett.124.051602}{\emph{Phys. Rev. Lett.}
  {\bfseries 124} (2020) 051602}
  [\href{https://arxiv.org/abs/1908.04325}{{\ttfamily 1908.04325}}].

\bibitem{Jensen:2010ga}
K.~Jensen, A.~Karch, D.~T. Son and E.~G. Thompson, \emph{{Holographic
  Berezinskii-Kosterlitz-Thouless Transitions}},
  \href{https://doi.org/10.1103/PhysRevLett.105.041601}{\emph{Phys. Rev. Lett.}
  {\bfseries 105} (2010) 041601}
  [\href{https://arxiv.org/abs/1002.3159}{{\ttfamily 1002.3159}}].

\bibitem{Iqbal:2010eh}
N.~Iqbal, H.~Liu, M.~Mezei and Q.~Si, \emph{{Quantum phase transitions in
  holographic models of magnetism and superconductors}},
  \href{https://doi.org/10.1103/PhysRevD.82.045002}{\emph{Phys. Rev.}
  {\bfseries D82} (2010) 045002}
  [\href{https://arxiv.org/abs/1003.0010}{{\ttfamily 1003.0010}}].

\bibitem{Pomarol:2019aae}
A.~Pomarol, O.~Pujolas and L.~Salas, \emph{{Holographic conformal transition
  and light scalars}},
  \href{https://doi.org/10.1007/JHEP10(2019)202}{\emph{JHEP} {\bfseries 10}
  (2019) 202} [\href{https://arxiv.org/abs/1905.02653}{{\ttfamily
  1905.02653}}].

\bibitem{Jarvinen:2011qe}
M.~Jarvinen and E.~Kiritsis, \emph{{Holographic Models for QCD in the Veneziano
  Limit}}, \href{https://doi.org/10.1007/JHEP03(2012)002}{\emph{JHEP}
  {\bfseries 03} (2012) 002} [\href{https://arxiv.org/abs/1112.1261}{{\ttfamily
  1112.1261}}].

\bibitem{Alanen:2010tg}
J.~Alanen, K.~Kajantie and K.~Tuominen, \emph{{Thermodynamics of Quasi
  Conformal Theories From Gauge/Gravity Duality}},
  \href{https://doi.org/10.1103/PhysRevD.82.055024}{\emph{Phys. Rev.}
  {\bfseries D82} (2010) 055024}
  [\href{https://arxiv.org/abs/1003.5499}{{\ttfamily 1003.5499}}].

\bibitem{Alanen:2011hh}
J.~Alanen, T.~Alho, K.~Kajantie and K.~Tuominen, \emph{{Mass spectrum and
  thermodynamics of quasi-conformal gauge theories from gauge/gravity
  duality}}, \href{https://doi.org/10.1103/PhysRevD.84.086007}{\emph{Phys.
  Rev.} {\bfseries D84} (2011) 086007}
  [\href{https://arxiv.org/abs/1107.3362}{{\ttfamily 1107.3362}}].

\bibitem{Anselmi:2000fu}
D.~Anselmi, L.~Girardello, M.~Porrati and A.~Zaffaroni, \emph{{A Note on the
  holographic beta and C functions}},
  \href{https://doi.org/10.1016/S0370-2693(00)00472-X}{\emph{Phys. Lett.}
  {\bfseries B481} (2000) 346}
  [\href{https://arxiv.org/abs/hep-th/0002066}{{\ttfamily hep-th/0002066}}].

\bibitem{Maldacena:1998im}
J.~M. Maldacena, \emph{{Wilson loops in large N field theories}},
  \href{https://doi.org/10.1103/PhysRevLett.80.4859}{\emph{Phys. Rev. Lett.}
  {\bfseries 80} (1998) 4859}
  [\href{https://arxiv.org/abs/hep-th/9803002}{{\ttfamily hep-th/9803002}}].

\bibitem{Rey:1998bq}
S.-J. Rey, S.~Theisen and J.-T. Yee, \emph{{Wilson-Polyakov loop at finite
  temperature in large N gauge theory and anti-de Sitter supergravity}},
  \href{https://doi.org/10.1016/S0550-3213(98)00471-4}{\emph{Nucl. Phys.}
  {\bfseries B527} (1998) 171}
  [\href{https://arxiv.org/abs/hep-th/9803135}{{\ttfamily hep-th/9803135}}].

\bibitem{Ryu:2006bv}
S.~Ryu and T.~Takayanagi, \emph{{Holographic derivation of entanglement entropy
  from AdS/CFT}},
  \href{https://doi.org/10.1103/PhysRevLett.96.181602}{\emph{Phys. Rev. Lett.}
  {\bfseries 96} (2006) 181602}
  [\href{https://arxiv.org/abs/hep-th/0603001}{{\ttfamily hep-th/0603001}}].

\bibitem{Freedman:1999gp}
D.~Z. Freedman, S.~S. Gubser, K.~Pilch and N.~P. Warner, \emph{{Renormalization
  group flows from holography supersymmetry and a c theorem}},
  \href{https://doi.org/10.4310/ATMP.1999.v3.n2.a7}{\emph{Adv. Theor. Math.
  Phys.} {\bfseries 3} (1999) 363}
  [\href{https://arxiv.org/abs/hep-th/9904017}{{\ttfamily hep-th/9904017}}].

\bibitem{Gubser:2008ny}
S.~S. Gubser and A.~Nellore, \emph{{Mimicking the QCD equation of state with a
  dual black hole}},
  \href{https://doi.org/10.1103/PhysRevD.78.086007}{\emph{Phys. Rev.}
  {\bfseries D78} (2008) 086007}
  [\href{https://arxiv.org/abs/0804.0434}{{\ttfamily 0804.0434}}].

\bibitem{Witten:2010zr}
E.~Witten, \emph{{A New Look At The Path Integral Of Quantum Mechanics}},
  \href{https://arxiv.org/abs/1009.6032}{{\ttfamily 1009.6032}}.

\bibitem{Behtash:2015zha}
A.~Behtash, G.~V. Dunne, T.~Sch{\"a}fer, T.~Sulejmanpasic and M.~{\"U}nsal,
  \emph{{Complexified path integrals, exact saddles and supersymmetry}},
  \href{https://doi.org/10.1103/PhysRevLett.116.011601}{\emph{Phys. Rev. Lett.}
  {\bfseries 116} (2016) 011601}
  [\href{https://arxiv.org/abs/1510.00978}{{\ttfamily 1510.00978}}].

\end{thebibliography}\endgroup

\end{document}